\documentclass[amsmath,amssymb,aps,prd,superscriptaddress,
reprint
]{revtex4-1}

\usepackage{empheq}
\usepackage{textcomp}
\usepackage{bigints}
\usepackage{colortbl}
\usepackage{multirow}
\usepackage{array}
\usepackage{graphicx}
\usepackage{braket}
\usepackage[table,xcdraw]{xcolor}
\usepackage[mathlines]{lineno}
\usepackage{float}
\usepackage{bigints}
\usepackage{makecell}
\usepackage{graphicx}
\usepackage{dcolumn}
\usepackage{bm}
\usepackage{hyperref}
\hypersetup{
    colorlinks = true,
    linkcolor = black,
    citecolor=blue, 
    }
\usepackage[capitalise]{cleveref}
\usepackage{siunitx}
\usepackage{graphicx} 
\usepackage{booktabs}
\usepackage{chemformula}
\usepackage[mathlines]{lineno}
\begin{document}

\title{Color symmetry breaking in a nonlinear optical microcavity}

\author{Luca O. Trinchão}
\thanks{These authors contributed equally to this work.}
\affiliation{Gleb Wataghin Institute of Physics, University of Campinas, Campinas, SP, Brazil}
\affiliation{Max Planck Institute for the Science of Light, Erlangen, Germany}

\author{Alekhya Ghosh}
\thanks{These authors contributed equally to this work.}
\affiliation{Max Planck Institute for the Science of Light, Erlangen, Germany}
\affiliation{Department of Physics, Friedrich Alexander University Erlangen-Nuremberg, Erlangen, Germany}

\author{Arghadeep Pal}
\thanks{These authors contributed equally to this work.}
\affiliation{Max Planck Institute for the Science of Light, Erlangen, Germany}
\affiliation{Department of Physics, Friedrich Alexander University Erlangen-Nuremberg, Erlangen, Germany}

\author{Haochen Yan}
\affiliation{Max Planck Institute for the Science of Light, Erlangen, Germany}
\affiliation{Department of Physics, Friedrich Alexander University Erlangen-Nuremberg, Erlangen, Germany}

\author{Toby Bi}
\affiliation{Max Planck Institute for the Science of Light, Erlangen, Germany}
\affiliation{Department of Physics, Friedrich Alexander University Erlangen-Nuremberg, Erlangen, Germany}

\author{Shuangyou Zhang}
\affiliation{Max Planck Institute for the Science of Light, Erlangen, Germany}
\affiliation{Department of Electrical and Photonics Engineering Technical University of Denmark Kgs. Lyngby, Denmark}

\author{Nathalia B. Tomazio}
\affiliation{Instituto de Física, Universidade de São Paulo, São Paulo, SP, Brazil}

\author{Flore K. Kunst}
\affiliation{Max Planck Institute for the Science of Light, Erlangen, Germany}
\affiliation{Department of Physics, Friedrich Alexander University Erlangen-Nuremberg, Erlangen, Germany}

\author{Lewis Hill}
\affiliation{Max Planck Institute for the Science of Light, Erlangen, Germany}

\author{Gustavo S. Wiederhecker}
\email[e-mail:]{ gsw@unicamp.br}
\affiliation{Gleb Wataghin Institute of Physics, University of Campinas, Campinas, SP, Brazil}

\author{Pascal Del’Haye}
\email[e-mail:]{ pascal.delhaye@mpl.mpg.de}
\affiliation{Max Planck Institute for the Science of Light, Erlangen, Germany}
\affiliation{Department of Physics, Friedrich Alexander University Erlangen-Nuremberg, Erlangen, Germany}

\date{\today}

\begin{abstract}
Spontaneous symmetry breaking leads to diverse phenomena across the natural sciences, from the Higgs mechanism in particle physics to superconductors and collective animal behavior.
In photonic systems, the symmetry of light states can be broken when two optical fields interact through the Kerr nonlinearity, as shown in early demonstrations with counterpropagating and cross-polarized modes.
Here, we report the first observation of color symmetry breaking in an integrated silicon nitride microring, where spontaneous power imbalance arises between optical mode at different wavelengths, mediated by the Kerr effect.
The threshold power for this effect is as low as \qty{19}{\milli\watt}.
By examining the system’s homogeneous states, we further demonstrate a Kerr-based nonlinear activation-function generator that produces sigmoid-, quadratic-, and leaky-ReLU-like responses. 
These findings reveal previously unexplored nonlinear dynamics in dual-pumped Kerr resonators and establish new pathways towards compact, all-optical neuromorphic circuits.
\end{abstract}

\maketitle

\section{Introduction}

Spontaneous symmetry breaking (SSB) refers to the transition of a system's state from being symmetric to asymmetric under an infinitesimal change of a control parameter. Signatures of SSB contribute to many areas of physics, spanning research on particle physics~\cite{baker1962spontaneous, higgs1964broken} and quantum field theory~\cite{adler1982einstein}, as well as plasmonics~\cite{barbillon2020applications} and liquid-crystal systems~\cite{dozov2001spontaneous, gibb2024spontaneous}. SSB also answers intriguing questions in the domains of Ising machines~\cite{mezard1992replica} and evolutionary biology~\cite{palmer2004symmetry}.

\begin{figure*}[]
\centering
\includegraphics[width=\textwidth]{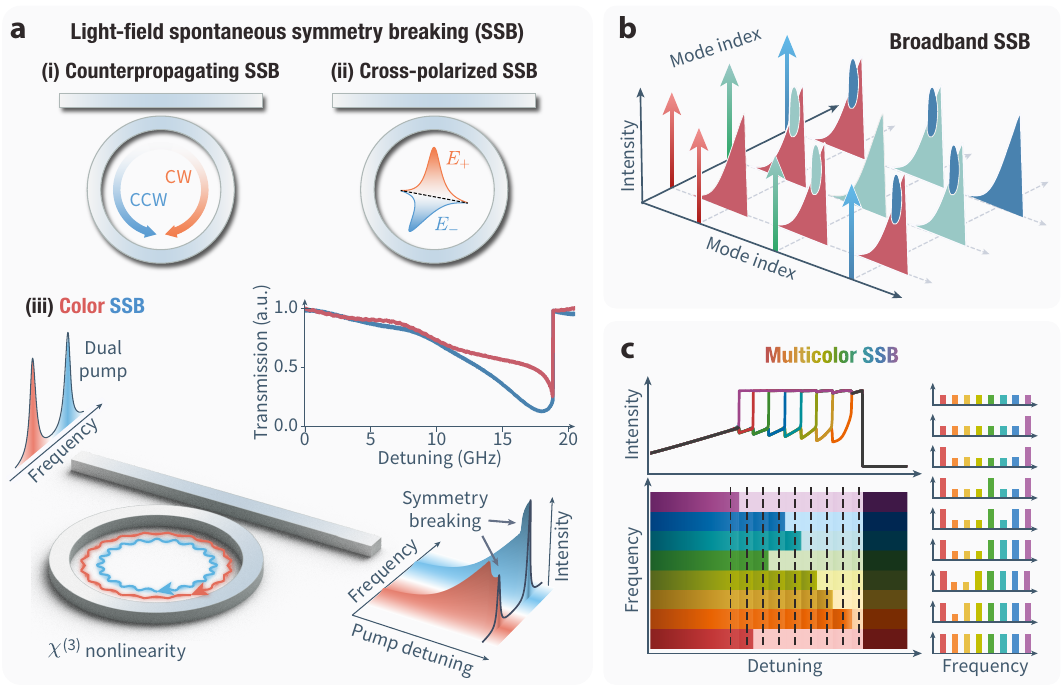}
\caption{\label{fig:1}
\textbf{(a)} \textit{Mechanisms of light-field spontaneous symmetry breaking (SSB) in Kerr microresonators}.
\textbf{(a.i)} Counterpropagating SSB between clockwise (CW) and counter-clockwise (CCW) modes.
\textbf{(a.ii)} Cross-polarized SSB between orthogonal polarization modes $E_+$ and $E_-$.
\textbf{(a.iii)} Color SSB between two modes at different frequencies, where dual pumping leads to an intracavity power imbalance. Inset: Experimental observation of color SSB. Red (blue) traces show the measured transmission of the low- (high-) frequency pumps when both lasers are simultaneously scanned across their respective resonances.
\textbf{(b)} \textit{Broadband SSB}: symmetry breaking can arise from the interaction of light in any pair of non-degenrate cavity resonances. No SSB occurs when both pumps excite the same resonance (diagonal terms).
\textbf{(c)} \textit{Multicolor SSB}: Numerical simulations of eight simultaneously pumped resonances showing high-dimensional SSB when the pump detunings are synchronously scanned, giving rise to a symmetry-broken comb of frequencies.
In the top-left panel, the black curve represents the symmetric solution branch. Once the SSB initiates, individual frequency modes (colored curves) undergo either an increase or a reduction in intracavity intensity, as depicted in the heatmap in the bottom panel.
The right panel displays the intracavity-intensity profiles across the spectrum, extracted at selected detuning values indicated by the dashed black lines in the colormap.
}
\end{figure*}

Within nonlinear photonics, light can exhibit spontaneous symmetry breaking, where the Kerr-nonlinearity drives an abrupt redistribution of intensities between identically driven resonant modes, yielding two equally likely outcomes in which one mode prevails over the other~\cite{del2017symmetry, hill2020effects, hill2025exceptional}. Recent research has associated the SSB of light fields in Kerr-cavities with the presence of exceptional points~\cite{hill2025exceptional,mazovasquez2025algebraic}, with experimental demonstrations in silica microrods~\cite{del2017symmetry, pal2024linear, ghosh2024phase, silver2021nonlinear, woodley2021self}, silicon nitride (Si$_3$N$_4$) microrings~\cite{zhang2025integrated}, Fabry-Perot cavities~\cite{moroney2022kerr}, and fiber ring resonators~\cite{garbin2020asymmetric, quinn2024coherent, lucas2025polarization, xu2021spontaneous}.

Previous research have extensively studied two main types of SSB of light fields in passive optical Kerr-cavities: \textit{(i)} chiral symmetry breaking between counterpropagating modes (\cref{fig:1}(a.i))~\cite{del2017symmetry, zhang2025integrated, pal2024linear, ghosh2024phase, anashkina2025phase, cao2020reconfigurable, pal2025microresonator, woodley2021self}; and \textit{(ii)} polarization symmetry breaking, where elliptically polarized light is generated from linearly polarized input
(\cref{fig:1}(a.ii))~\cite{garbin2020asymmetric, quinn2023random, quinn2024coherent, moroney2022kerr, hill2024symmetry, campbell2024frequency}. Recent efforts have focused on extending SSB from two-level systems to multi-level systems by combining chiral and polarization symmetry breaking in a single resonator~\cite{hill2023multi} and studying networks of coupled resonators~\cite{pal2024linear, ghosh2024controlled, ghosh2023four}. 
While homogeneous solutions (constant field intensity throughout the resonator at any instance) may suffice to demonstrate SSB in some systems~\cite{mai2024spontaneous, bitha2023bifurcation, mazovasquez2025algebraic}, intriguing symmetry breaking dynamics of temporal structures such as bright and dark solitons~\cite{xu2021spontaneous, ghosh2025spontaneous, xu2022breathing, hill2024symmetry, campbell2025vectorial, campbell2024frequency} and polarization faticons~\cite{lucas2025polarization, hill2025controlling} have also been unveiled.
Beyond their fundamental significance, symmetry-broken optical states enable all-optical functionalities such as isolators~\cite{del2018microresonator, white2023integrated}, memories~\cite{zhang2025integrated}, gyroscopes~\cite{silver2021nonlinear}, switches, and universal logic gates~\cite{ghosh2024phase}, as well as random number generation~\cite{quinn2023random} and coherent Ising machines~\cite{quinn2024coherent}.

For our experiments, we use silicon nitride (Si$_3$N$_4$) microresonators.
Si$_3$N$_4$ has become a leading platform for integrated photonics due to its CMOS-compatible fabrication techniques, low propagation loss, and broad transparency window~\cite{xiang2022silicon, zhang2024low}. Its high Kerr-nonlinearity has enabled observation of different nonlinear optical interactions, such as the formation of frequency combs~\cite{okawachi2011octave, pal2025hybrid}, bright~\cite{ji2025deterministic} and dark~\cite{zhang2023geometry} solitons, and spontaneous symmetry breaking~\cite{zhang2025integrated}, feasible.

In this work, we report on the observation of \emph{color symmetry breaking} (\cref{fig:1}(a.iii)), where SSB emerges between two nondegenerate frequency modes in an integrated Si$_3$N$_4$ Kerr microresonator.
By coherently driving distinct resonances with a bi-chromatic pump, we find a spontaneous imbalance in the optical power of the two modes (\cref{fig:1}(a.iii, inset)).
Color SSB fundamentally differs from prior counterpropagating and cross-polarized demonstrations, which are limited to a two-dimensional space (propagation direction or polarization).
In contrast, color SSB can be carried out with any combination of resonant modes of the system (\cref{fig:1}(b)).
This unlocks an extra degree of freedom in \emph{frequency} for on-chip photonic symmetry breaking. 
This capability marks a transition from monochromatic to broadband symmetry-breaking photonics, paving the way for synthetic, high-dimensional multicolor SSB across multiple optical frequencies (\cref{fig:1}(c)).
Although recent works have uncovered intriguing fast-time dynamics in resonators pumped with multiple lasers of different frequencies~\cite{zhang2019sub, zhang2020spectral, moille2023kerr, moille2024parametrically, tomazio2024tunable}, we focus here on the rich homogeneous solution states of the system.

\begin{figure*}[]
\centering
\includegraphics[width=0.95\textwidth]{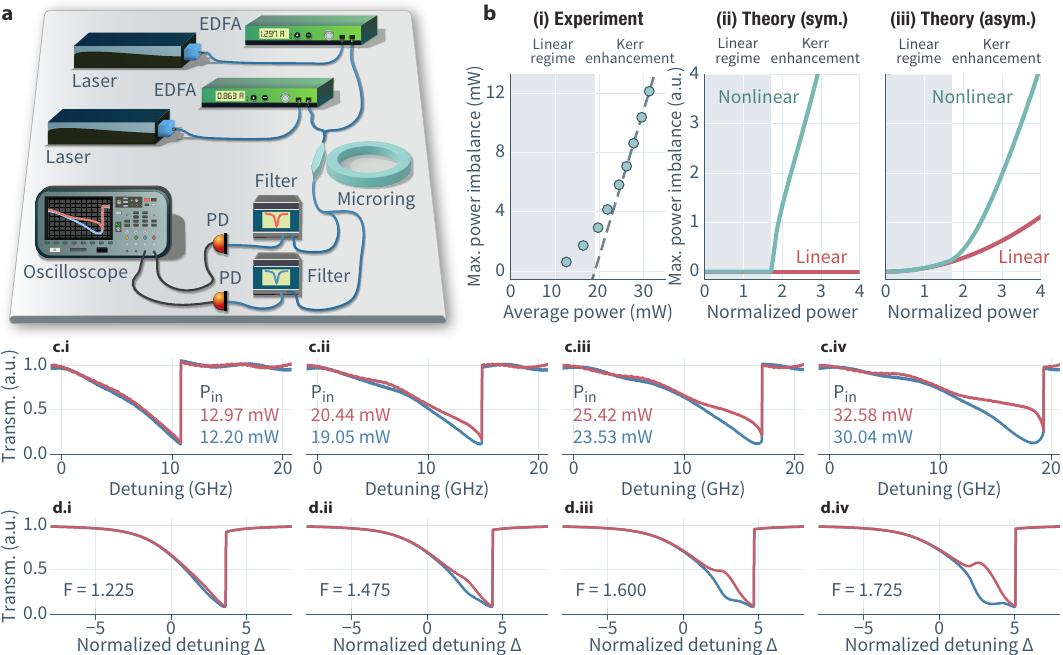}
\caption{\label{fig:2}
\textbf{(a)} Schematic of the experimental setup. EDFA: erbium-doped fiber amplifier, PD: photodetector.
\textbf{(b)} Color symmetry breaking threshold. Experimental (i) and theoretical (ii, iii) power imbalance between the two propagating modes ($a_\mathrm{blue}$ corresponding to a pump at \qty{1571}{\nano\meter} and $a_\mathrm{red}$ corresponding to a pump at \qty{1577}{\nano\meter}) as a function of input power. The shaded region marks the regime dominated by the cavity’s linear response, while the white region highlights the region of Kerr-induced rapid enhancement of power imbalance. In (i), scattered data points are extracted from the measured transmission, and the dashed line indicates the linear asymptote of the high-power response. In (ii) and (iii), numerical results for the symmetric and asymmetric models, respectively, illustrate the contributions from the linear response alone and from Kerr-induced XPM. \textbf{(c)} Experimental color symmetry breaking traces as a function of detuning for increasing optical power coupled into the bus waveguide ($P_{in})$ (i–iv), as indicated. The intensity of symmetry breaking increases with power.
\textbf{(d)} Numerical reproduction of (c) based on the asymmetric model. The normalized input powers $F$ are shown in the plots (see Methods).
}
\end{figure*}

Furthermore, by investigating the asymmetries in coupling and losses between the two pumped modes, we show that they act as biases in the otherwise random symmetry-breaking process~\cite{garbin2020asymmetric, mazovasquez2025algebraic}, yet the system retains SSB-like behavior governed by the unfolding of a characteristic pitchfork bifurcation~\cite{golubitsky1979analysis}.
An example of an application of this type of symmetry breaking could be used an on-chip programmable nonlinear function generator, implementing multiple activation functions across the whole electromagnetic spectrum.
\section{Results and discussion}
\subsection{Experimental observations}

We consider a Kerr microring resonator coherently driven by two pumps simultaneously tuned across two spectrally separated resonances of the cavity, as illustrated in \cref{fig:1}(a.iii).
As both laser frequencies approach resonance, intracavity power builds up, and the two modes interact through the Kerr nonlinearity~\cite{trinchão2025mapping}.
This coupling occurs via cross-phase modulation (XPM), which shifts the resonance frequency experienced by each pump proportionally to the intensity of the other.
Because XPM is twice as strong as self-phase modulation (SPM), small power fluctuations between the two modes are unequally self-amplified.
Above a certain power threshold, more intense light gets coupled to one of the cavity modes (say, blue), while the other resonance (resonance red) is driven away due to XPM from its corresponding pump (pump red), giving rise to SSB~\cite{hill2020effects, woodley2018universal}.
This is highlighted in inset of Fig.~\ref{fig:1}a(i,ii).
In an ideally symmetric system (resonances with equal coupling conditions, losses and identical pump conditions), this imbalance arises spontaneously, with the dominant mode selected at random.

Figure~\ref{fig:2}(a) shows the experimental setup used to explore color symmetry breaking in an integrated Si$_3$N$_4$ resonator.
Figure \ref{fig:2}(b.i) shows the measured transmitted power imbalance between the two lasers as a function of the on-chip power in the bus waveguide.
The imbalance appears at input powers around \qty{12}{\milli\watt} and grows approximately linearly at higher powers.
Two distinct contributions can be identified.
At low powers, imbalance arises from intrinsic asymmetries in coupling and losses (and consequently in the loaded $Q$ factors), which are responsible for the cavity's linear response. The resonance with the narrower linewidth accumulates energy more rapidly. At higher powers, an additional contribution emerges from Kerr-effect-induced XPM~\cite{woodley2018universal, hill2020effects}, which we refer to as \emph{Kerr asymmetry enhancement}.
Here, the stronger optical mode shifts the partner resonance further away from the pump, enhancing the imbalance and producing transmission features of symmetry-broken states (\cref{fig:2}(c)).

As mentioned earlier, the intrinsic asymmetry of the two resonances ($Q_{blue}=5.3\times10^5$ and $Q_\mathrm{red}=5.5\times10^5$, nearly critically coupled) leads to a mismatch in the coupled powers, which appears even below the SSB threshold.
While imbalance becomes observable around \qty{12}{\milli\watt}, the calculated threshold for an ideal symmetric system is \qty{18}{\milli\watt}.
By extrapolating the high-power asymptote, a zero crossing appears near \qty{19}{\milli\watt}, indicating the true onset of Kerr asymmetry enhancement.
Figures \ref{fig:2}(b.ii) and \ref{fig:2}(b.iii) present the corresponding theoretical predictions for symmetric and asymmetric systems, respectively. In the symmetric case, a sharp threshold is observed, marked by the abrupt onset of imbalance due to SSB of the optical modes. In contrast, the asymmetric case exhibits a finite imbalance already in the linear regime; beyond a critical threshold, this imbalance is rapidly amplified by XPM.

Starting from a low-power symmetric transmission for both pumps (\cref{fig:2}(c.i)), increasing the input power results in a characteristic bubble-shaped imbalance between the transmitted fields (\cref{fig:2}(c.ii)).
This feature marks the onset of Kerr-induced symmetry breaking, which intensifies with power (\cref{fig:2}(c.iii and iv)).
These observations agree with our theoretical studies based on mean-field coupled equations (\cref{fig:2}(d), see \cref{sec:Theory}).
Notably, the system exhibits a consistent bias favoring the blue pump as the dominant intracavity mode, indicating the presence of intrinsic asymmetries, discussed in detail below.

\subsection{Theory of color SSB} \label{sec:Theory}

\begin{figure}[t!]
\centering
\includegraphics[width=0.95\linewidth]{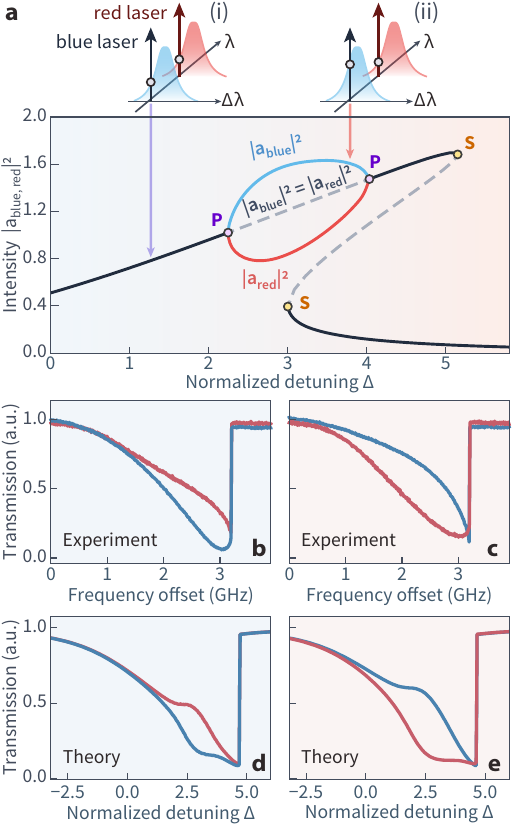}
\caption{\label{fig:3}
\textbf{(a)} Bifurcation diagram of color symmetry breaking. Intracavity power is plotted as a function of the laser detuning. Black: symmetric solutions. Red and blue: symmetry-broken solutions, where either $a_\mathrm{red}$ or $a_\mathrm{blue}$ dominates. P: pitchfork bifurcation. S: saddle-node bifurcation. Solid (dashed) lines denote stable (unstable) branches. Insets: (i) Symmetric regime, where both lasers are equally detuned from their respective resonances; (ii) symmetry-broken regime, where one resonance is pulled closer to its laser while the other is shifted away.
\textbf{(b, c)} Experimental observation of color symmetry breaking, showing deterministic flipping of the dominant mode when the red laser frequency is reduced by \qty{195}{\mega\hertz}.
\textbf{(d, e)} Numerical reproductions of (b) and (c) based on the asymmetric model (see Methods).
}
\end{figure}

The slow-time evolution of the intracavity modes can be described with a mean-field approximation~\cite{chembo2010modal, lugiato1987spatial, hill2025exceptional}:
\begin{equation}
    \frac{\partial a_{\mathrm{c}}}{\partial t} = - \left(1 + i\Delta\right) a_{\mathrm{c}} + i \left(|a_\mathrm{c}|^2 + 2\,|a_\mathrm{c'}|^2 \right) a_\mathrm{c} + \sqrt{F},
    \label{CME_symmetric}
\end{equation}
where $\mathrm{c},\mathrm{c'}\in\{\mathrm{red}, \mathrm{blue}\}$ and $\mathrm{c} \neq\mathrm{c'}$.
Here, $a_\mathrm{red}$ and $a_\mathrm{blue}$ represent the normalized mode amplitudes of the low- and high-frequency resonant modes, respectively (see Methods).
The parameters $\Delta$ and $F$ denote the normalized detuning and driving power, which are assumed identical for both modes.
The nonlinear terms account for SPM and XPM effects, with XPM being twice as strong as SPM and responsible for coupling between the two modes.

At a certain detuning threshold, the system undergoes a pitchfork bifurcation (\cref{fig:3}(a)), giving rise to two mirrored asymmetric states corresponding to dominance of either optical mode~\cite{del2017symmetry, hill2020effects, hill2025exceptional}.

\begin{figure}[]
\centering
\includegraphics[width=\linewidth]{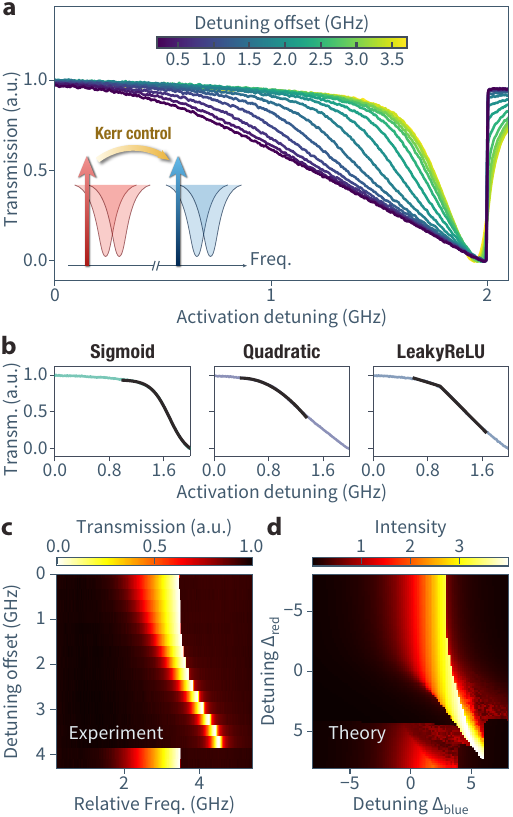}
\caption{\label{fig:4}
\textbf{(a)} Experimental transmission traces of the probe laser show a variety of nonlinear responses as a function of the detuning. Different colors correspond to responses obtained by changing the control laser detuning. 
For clarity, the detuning axis of each trace is offset so that the end of the optical-bistability region aligns across all curves, highlighting the distinct nonlinear behaviors. 
The inset illustrates the Kerr-mediated interaction through which the control laser modifies the probe laser response.
\textbf{(b)} Examples of selected responses fitted with Sigmoid, Quadratic, and LeakyReLU-type functions.
\textbf{(c, d)} Experimental (c) and numerical (d) heatmaps of the probe laser transmission as a function of its own relative detuning and the control laser detuning offset.
}
\end{figure}

When asymmetries (such as unequal coupling, detuning, or loss rates) are introduced, the pitchfork bifurcation turns into a more gradual separation of the powers in the two modes~\cite{garbin2020asymmetric, golubitsky1979analysis, mazovasquez2025algebraic} while the system becomes deterministic: the mode with the higher loaded $Q$ or smaller detuning consistently accumulates more energy, defining the dominant branch.
We experimentally demonstrate this controlled bias by tuning the relative detuning offset between the two pumps to favor either the red or blue mode (\cref{fig:3}(b,c)).
When the red-pump frequency is decreased by only \qty{195}{\mega\hertz}, the dominant state consistently flips in its favor.
This observation is supported by numerical investigations that account for asymmetries (see Methods and \cref{fig:3}(c,d)).
Similar approaches have previously demonstrated that multiple asymmetries can be engineered to compensate one another and restore effective symmetry~\cite{garbin2020asymmetric, mazovasquez2025algebraic}.
However, the intrinsic difference in the quality factors of the two resonances prevents restoring effective symmetry throughout the entire detuning scan.

\subsection{Kerr-driven activation functions}

In this section, we demonstrate how the inter-modal Kerr-interaction responsible for SSB, discussed in the previous sections, can enable all-optical nonlinear function generation. Machine learning is useful for classifications and predictions across a wide range of technologies and scientific disciplines~\cite{jordan2015machine, krizhevsky2012imagenet}, while it is pushing the limits of digital electronic architectures in terms of complexity and power consumption~\cite{Thompson2023Computational}. Optical neural networks, on the other hand, have recently positioned themselves as tools to address these challenges~\cite{lin2018all, feldmann2021parallel, volpato20251t, slinkov2025all}.

However, most of the available research relies on opto-electronic nonlinearities to implement the nonlinear activation function~\cite{williamson2019reprogrammable, bandyopadhyay2024single}, which limits the processing speed. 

In this work we propose the use of the Kerr-nonlinearity for on-chip function generation.
In the previous sections, we examined the slow-time homogeneous solutions of dual-pumped microresonators under symmetric pumping conditions, highlighting the emergence of SSB. To demonstrate nonlinear function generation, we now turn attention to asymmetric pumping configurations, in which one pump (the control) is held at a fixed detuning while the wavelength of the second pump (the probe) is scanned.

Figure~\ref{fig:4}(a) shows that the presence of a resonant control laser can be used to tailor the nonlinear response in the transmission of the probe laser, mediated by the joint contributions of Kerr XPM and thermal effects.
When the wavelength of the probe laser is scanned, its transmission exhibits diverse nonlinear shapes that can be continuously tuned by adjusting the control laser wavelength.
As shown in \cref{fig:4}(b), different responses can be fitted by sigmoid, quadratic, and LeakyReLU-type functions, which are required by the machine learning community for classification tasks~\cite{dubey2022activation}.
Moreover, because XPM is a broadband effect capable of coupling the dynamics of many cavity modes, this approach naturally supports wavelength-division multiplexing, enabling parallel information encoding across multiple frequency channels over the entire electromagnetic spectrum~\cite{slinkov2025all}.

\cref{fig:4}(c,d) present experimental and numerical heatmaps of the probe laser transmission as a function of its own relative detuning and the detuning offset of the control pump, effectively mapping the $\left(\Delta_\mathrm{blue}, \Delta_\mathrm{red}\right)$ phase diagram. 
Although optical broadening is observed experimentally due to the combined Kerr and thermal phase shifts~\cite{trinchão2025mapping}, whereas the numerical model includes only Kerr nonlinearities, the experiments and the simulations show excellent agreement.
When the control laser approaches its resonance, it progressively reduces the Kerr-(and thermal) broadening of the probe resonance. 
This type of auxiliary-pump–assisted narrowing has previously been exploited in the context of thermal stabilization of temporal solitons~\cite{zhang2019sub}. 
However, the close agreement between our measurements and Kerr-only simulations demonstrates that this response also purely arises from faster Kerr-mediated dynamics, even in the absence of thermal effects.
This indicates that Kerr-based activation functions in our system can operate at very high speeds, limited only by the cavity buildup time~\cite{trinchão2025mapping}.

In summary, we have investigated the rich homogeneous solutions of bichromatically pumped Kerr microresonators.
Under symmetric conditions, we observe the emergence of color symmetry breaking, establishing new strategies for symmetry breaking phenomena in photonic platforms.
We further explore  the system under asymmetric pumping, revealing diverse nonlinear responses that can be tuned for implementing Kerr-based on-chip activation functions.
Our results show the prospects of bichromatic interactions as versatile nonlinear platforms for realizing controlled, integrated photonic neuromorphic computing.

\section{Methods}

\subsection{Experimental details}

The experimental concept is shown in \cref{fig:2}(a).
Two resonances of the same mode family at $\lambda_\mathrm{red} = \qty{1577}{\nano\meter}$ and $\lambda_\mathrm{blue} = \qty{1571}{\nano\meter}$, separated by three free spectral ranges, are simultaneously scanned using two independently tunable lasers.
Both lasers are driven by a common ramp signal to ensure synchronous scanning at identical detuning.
Residual detuning offsets can be dynamically corrected by adjusting the central frequency of each laser scan.
Each laser is amplified and sent through a variable optical attenuator (VOA) to control the relative pump powers.
A second VOA placed after recombination regulates the total power injected in the microresonator.
Light is coupled into and out of the chip using lensed fibers, and the transmitted signals are spectrally separated by optical filters and recorded with an oscilloscope.

\subsection{Unbalaced theory model}
As mentioned in the main text, the quality factors of the resonances, spectrally separated by certain FSRs, differ. Therefore, to simulate identical input conditions, a small power mismatch and relative detuning are introduced between the two lasers. To model this, we use a modified version of Eq.~\eqref{CME_symmetric}:
\begin{subequations}
\begin{align}
    &\frac{\partial a_{r}}{\partial t} = - \left(1 + i\Delta_r\right) a_{r} + i \left(|a_r|^2 + 2\,|a_b|^2 \right) a_r + \sqrt{F_r},\\
    &\frac{\partial a_{b}}{\partial t} = - \left(\eta + i\eta\Delta_b\right) a_{b} + i \left(\eta|a_b|^2 + 2\,|a_r|^2 \right) a_b + \eta\sqrt{F_b},
    \label{CME_asymmetric}
\end{align}
\end{subequations}
where $a_c=\sqrt{\left(2/\kappa_{c}\right)}E_c$, $a_c$ ($E_c$) is the normalized (unnormalized) electric field envelop for $c\in\{r,b\}$. Here $r~(b)$ denotes the red~(blue) mode, corresponding to the higher~(lower) wavelength component. The internal and external losses of the resonance with index $c$, denoted by $\kappa_{c,l}$ and $\kappa_{c,\text{ex}}$, respectively, constitute the total loss of the system $\kappa_c$. We consider equal coupling losses for the two resonances, $\kappa_{r,\text{ex}}=\kappa_{b,\text{ex}}=\kappa_{\text{ex}}$, and define $\eta=\left(\kappa_r/\kappa_b\right)$. The normalized input powers are given as $\sqrt{F_c}=\sqrt{\frac{8\kappa_{\text{ex}}}{\kappa^3_c}}s_{\text{in},c}$, where $s_{\text{in},c}$ is the unnormalized input-field amplitude. The normalized detunings is defined as $\Delta_c=\frac{2(\omega_{c,0}-\omega_{c})}{\kappa_c}$, where, $\omega_c$ is the laser frequency and $\omega_{c,0}$ is the corresponding nearest resonance frequency.

\subsection{Multicolor SSB}

The Kerr XPM effect nonlinearly couples the intracavity dynamics of all resonant modes. Their homogeneous evolution is governed by~\cite{chembo2010modal}:
\begin{equation}
\label{CME_manymodes}
    \frac{\partial a_{j}}{\partial t} = - \left(1 + i\Delta\right) a_{j} + i \left(|a_j|^2 + 2\,\sum_{k \neq j}|a_k|^2 \right) a_{j} + \sqrt{F}.
\end{equation}
Here, $j,k$ label the cavity modes, and frequency-mixing terms are neglected.
This approximation is valid when operating at power levels below the threshold for generating new frequencies via four-wave mixing (FWM), or when dispersion engineering~\cite{pal2023machine} suppresses the phase-matching of FWM interactions.\\
\indent \cref{fig:1}(c) shows numerical simulations of eight symmetrically pumped modes for a normalized drive $F^2 = 1.4$. As the detuning is swept, seven symmetry-breaking thresholds appear, at which individual modes randomly transition between the upper and lower intensity branches.\\


\section*{Acknowledgements}
We thank Dr. George Ghalanos and Prof. Yaojing Zhang for initial discussions.\\
\indent LOT, AG, and AP contributed equally to the work. LOT and AP did the experiment with help from AG. AG and LOT did the simulations. AG, HCY and AP fabricated the samples. LOT, AG, AP, PDH, and GSW wrote the manuscript with the help of all other authors. PDH and GSW supervised the project.\\
\indent This work was funded by the European Union’s H2020 ERC Starting Grant ``CounterLight” 756966, MQV Project TeQSiC, the German Federal Ministry of Research, Technology and Space, Quantum Systems, 13N17314, 13N17342, the Max Planck Society, and the Max Planck School of Photonics.
This work was supported by São Paulo Research Foundation (FAPESP) through grants 
18/15580-6, 
18/25339-4, 
21/10334-0, 
23/09412-1, 
24/15935-0, 
25/04049-1, 
20/04686-8, 
and Coordenação de Aperfeiçoamento de Pessoal de Nível Superior - Brasil (CAPES) (Finance Code 001).
SZ acknowledge support from Deutsche Forschungsgemeinschaft project 541267874.

\bibliography{bib.bib}

\end{document}